\documentclass[journal]{IEEEtran}
\ifCLASSINFOpdf
\else
\fi
\usepackage{bbm}
\usepackage{fancyvrb}
\usepackage{algorithm}
\usepackage{algorithmicx}
\usepackage[noend]{algpseudocode}

\usepackage{mynotes}

\begin{document}
\title{Computer Assisted Localization of a Heart Arrhythmia}

\author{Chris~Vogl,
	Peng~Zheng,
        Stephen P.~Seslar,
        and~Aleksandr Y. Aravkin% <-this % stops a space
\thanks{C. Vogl is with the Lawrence Livermore National Laboratory}% <-this % stops a space
\thanks{P. Zheng and A. Aravkin are with the Department of Applied Mathematics at the University of Washington}% <-this % stops a space
\thanks{S. Seslar is with Seattle Children's Pediatric Hospital and the University of Washington}
\thanks{Manuscript received ***}}


\markboth{Preprint}%
{Shell \MakeLowercase{\textit{et al.}}: Optimal Localization of a Heart Arrhythmia }



\maketitle

\begin{abstract}
We consider the problem of locating a point-source heart arrhythmia 
using data from a standard diagnostic procedure, where a reference 
catheter is placed in the heart, and arrival times from 
a second diagnostic catheter are recorded as the diagnostic catheter
moves around within the heart. 

We model this situation as a nonconvex feasibility problem, where given 
a set of arrival times, we look for a source location that is consistent
with the available data. We develop a new optimization approach 
and fast  algorithm to obtain online proposals for the next location 
to suggest to the operator as she collects data. 

We validate the procedure using a Monte Carlo simulation based on patients' electrophysiological data.
The proposed procedure robustly and quickly locates the source of arrhythmias without 
any prior knowledge of heart anatomy.

\end{abstract}


% \fi
\IEEEpeerreviewmaketitle

\section{Introduction}

Catheter ablation is the treatment of choice to diagnose and treat cardiac arrhythmias. 
Accurately determining the origin of a cardiac arrhythmia is of critical importance in catheter ablation procedures.  
In many instances, arrhythmias originate from a focal point source and the electrical signal spreads concentrically in all directions away from that point. The traditional method of localizing such arrhythmias involves a user-directed movement of a mapping catheter through a cardiac chamber ---
 a somewhat haphazard ``hunt and peck'' method. 
The arrhythmia signal arrival time on the roving map catheter is compared against the signal arrival time on a stationary reference catheter until the {\it earliest relative timing} site is identified.  \\
Based on inherent limitations in how humans recognize patterns, this process requires a significant area of the heart chamber to be mapped before the operator can start to hone in on the location of the arrhythmia source. The operator is essentially required to solve an optimization problem by hand
to minimize the arrival time relative to that obtained by a stationary catheter.  \\
We propose a computer-assisted mapping system that can effectively use all available information
and point the operator to the `next touch' location.  This approach can substantially reduce the time and number of touch points needed to reliably determine the site of arrhythmia origin. We develop a method for arrhythmia localization and evaluate it using Monte Carlo simulations based on a deidentified electroanatomic map from a patients that underwent mapping and acutely successful ablation of a focal arrhythmia using the Rhythmia mapping system\footnote{Boston Scientific,  Marlborough, MA, USA.}.\\
{\bf Related work.} We found one prior approach to localizing arrhythmia using optimization~\cite{weber2017novel}. 
The approach is similar in principle, and uses optimization to recommend the next point to sample by the operator. 
The key differentiating factor is that~\cite{weber2017novel} assumes the existence of an anatomical map; the approach is predicated on being able to solve a linear regression problem that uses {\it every available nodal point} as a potential origin, and then pick the one that
best fits the available data. In contrast, we make no assumptions about the existence of anatomical maps; 
our proposals are based only on observations taken by the operator, and we can help patients who have undergone no prior mapping studies. 
In addition, the algorithm of~\cite{weber2017novel}, for each recommendation, must solve a number of regression problems 
equal to the number of points in their mesh. In contrast, we only need to solve {one feasibility problem} to find the next point proposal. \\
The paper proceeds as follows. In Section~\ref{sec:formulation}, we formulate 
arrhythmia localization as a feasibility problem and derive an associated optimization problem. 
In Section~\ref{sec:algorithm} we develop a fast algorithm, 
with practical considerations for dealing with noisy data and online application of arrhythmia localization. 
In Section~\ref{sec:results} we present 
results of the approach using patient arrival time data, and end with conclusions in Section~\ref{sec:conclusion}.

\section{Formulation}
\label{sec:formulation}

We are given a set of (ordered) relative arrival times 
\begin{equation}
\label{eq:timings}
0=t_1\le \ldots\le t_m,
\end{equation}
obtained by finding the differences between times recorded using a reference 
catheter and a diagnostic catheter at 3D locations $ x_1,\ldots, x_m$ within the heart.
{\bf Simple Feasible Region.}
We look for a source location $x_s$ consistent with available  observations. 
We assume that the local signal transmission speed $s$ around the source $x_s$ is known and given.
Denoting the actual arrival times by $\hat t_i$, we get  relations
\[
\hat t_i \ge t_i, \quad i = 1, \ldots, m,
\]
which translate to the constraints
\begin{equation}
\label{eq:constraints}
\|x_s - x_i\|_2 \ge s t_i, \quad i = 1, \ldots, m.
\end{equation}

\begin{figure}
\center
\begin{tabular}{cc}
\includegraphics[width=0.46\linewidth]{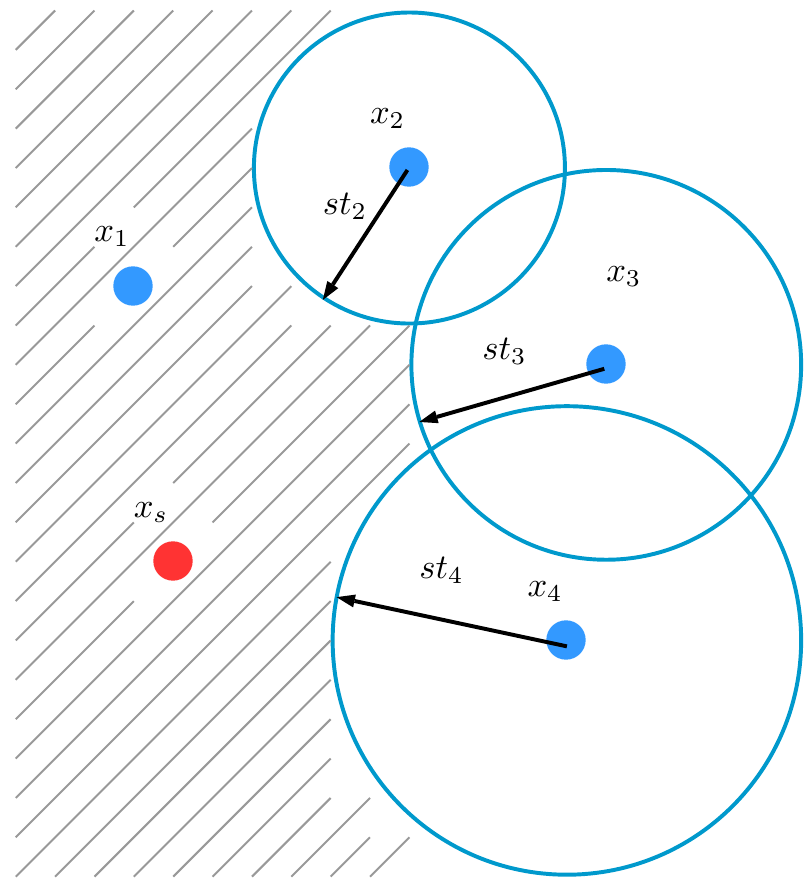}
&\includegraphics[width=0.46\linewidth]{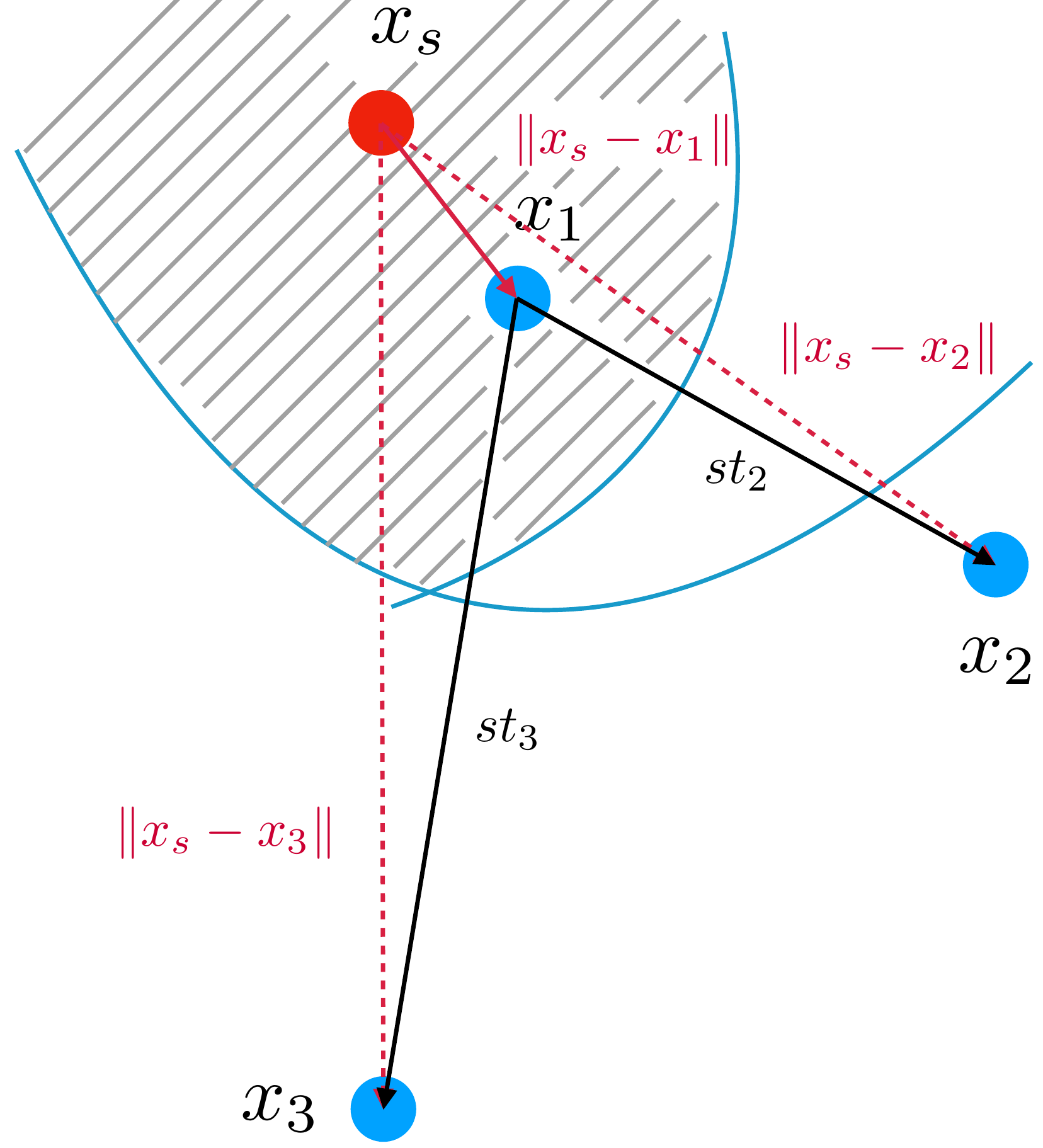}\\
(a) & (b)
\end{tabular}
\caption{\label{fig:geom}Simple timing relationships define a feasibility problem for locating the source location $x_s$ of the arrhythmia. 
(a) The set $\Omega_1$ described by~\eqref{eq:constraints}. (b) The set $\Omega_2$ described by~\eqref{eq:coupled}.}
\end{figure}

Finding a source that is consistent with the available observations
is equivalent to finding $x_s$ that satisfies \eqref{eq:constraints}. 
The feasible region from~\eqref{eq:constraints} is shown in panel (a) of Figure~\ref{fig:geom}; 
the true source must lie {outside} of the union of the disks shown in the figure.

{\bf Coupled Feasible Region.} 
A more powerful formulation incorporates first arrival information, similar to the fast 
marching method used to solve the Eikonal equation~\cite{sethian1996fast}. 
We look for $x_s$ satisfying
\begin{equation}
\label{eq:coupled}
\| x_s - x_i\| \ge \| x_s-x_1\| + s t_i,\; i \geq 2.
\end{equation}
%\begin{equation}
%\label{eq:obj}
%\begin{aligned}
%\min_{ x_s, \{w_i\}}~~&\frac{1}{2}\sum_{i=1}^m\| x_s - x_i\|^2,\\
%\text{s.t.}~\| x_s& - x_i\| \ge \| x_s-x_1\| + s t_i,\; i \geq 2,\end{aligned}
%\end{equation}
%where we use a reverse triangle inequality: a path between $x_i$ for $i \geq 2$ and $x_s$
%should take longer than moving from $x_s$ to $x_1$, and then from $x_s$ to $x_i$. 

The fast marching method and the inequalities~\eqref{eq:coupled} enforce a reverse triangle inequality: 
the time the signal takes to go from $x_s$ to $x_i$ is is larger than the time needed to go from $x_s$ to $x_1$ and then $x_1$ to $x_i$;
otherwise we would not have observed the given distribution of arrival times.  
The resulting region is shown in panel (b) of Figure~\ref{fig:geom}.

\section{Algorithm}
\label{sec:algorithm}

Nonconvex feasibility problems can be solved using optimization techniques such as
alternating optimization methods or Douglas-Rachford splitting~\cite{li2016douglas}. 
These algorithms can take hundreds to thousands of iterations for simple problem instances~\cite[Table 1]{li2016douglas}.
We propose a new relaxation that converges very rapidly, generating a feasible solution $x_s$ 
within a few iterations in most instances.

In developing relaxations for~\eqref{eq:constraints} and~\eqref{eq:coupled}, 
we use the ideas recently developed by~\cite{zheng2018fast}.
%The approach for solving~\eqref{eq:constraints} is very similar and so omitted here. 
We introduce auxiliary variables $w_i$ to approximate each $x_s - x_i$, 
and minimize over both $x_s$ and these auxiliary variables.
\begin{equation}
\label{eq:obj}
\begin{aligned}
\min_{ x_s, w}~~& f(x_s, w) :=\frac{1}{2}\sum_{i=1}^m\| x_s -  x_i -  w_i\|^2\\
\text{s.t.}~~&w \in \Omega,
\end{aligned}
\end{equation}
with $\Omega$  a special set described below. 
The original sets defined by~\eqref{eq:constraints} and~\eqref{eq:obj} are both subset of $\mathbb{R}^3$.
The set describing~\eqref{eq:constraints} is a simple subset of $\mathbb{R}^{3m}$: 
\[
\Omega_1 := \bigoplus_{i=1}^m\{w_i : \| w_i\|_2 \ge s t_i\}.
\]
The set in~\eqref{eq:coupled} is also a subset of $\mathbb{R}^{3m}$, with more complex structure:
\[\begin{aligned}
\Omega_2 := &\{w_1: w_1\in\mathbb{R}^3\}\oplus\\
&\bigoplus_{i=2}^m\{w_i:  \| w_i\|_2 \ge \|w_1\|_2 +s t_i\},
\end{aligned}\]
where $\oplus$ denote the direct sum.
The original problem for calculating a projection onto a nonconvex set is difficult.
By relaxing the formulation, the objective becomes more tractable, yielding a simple update rule
using the structure of $\Omega$. We 
also have a guarantee of optimality for the original feasibility problem 
based on the objective value of~\eqref{eq:obj}:
\begin{itemize}
\item Any $x_s$ satisfying~\eqref{eq:constraints} gives a global minimizer of~\eqref{eq:obj} with objective value $0$, 
by $\overline w_i = x_s - x_i$. 
\item Any solution with zero objective value gives $\overline x_s$ feasible with respect to~\eqref{eq:constraints} or~\eqref{eq:coupled}.
\end{itemize}

Problem~\eqref{eq:obj} may have a nonzero optimal value, in which case the `relaxed' solution $x_s$ 
will not satisfy the original formulation. However, in practice we find a feasible point in each iteration. 
To solve~\eqref{eq:obj}, we minimize over $x_s$ and $w_i$. Given $\{w_i\}$, 
we have a closed for solution for $x_s$:
%In this case, we partially minimize over $x_s$. In fact, 
%we have a simple closed form solution for $x_s$ as a function of $w_i$:
\begin{equation}
\label{eq:xupdate}
x_s = \frac{1}{m} \sum_{i=1}^m (x_i + w_i). 
\end{equation}
To find $\{w_i\}$ given $x_s$, we have to solve
\[
\min_{w\in\Omega} 0.5\sum_{i=1}^m\|w_i - (x_s - x_i)\|_2^2.
\]
When $\Omega = \Omega_1$, we have a closed form solution for the projection problem:
\begin{equation}
\label{eq:wupdate}
w_i = \frac{x_s - x_i}{\|x_s - x_i\|} \max(st_i, \|x_s - x_i\|_2).
\end{equation}
When $\Omega = \Omega_2$, the projection is found by solving 
\[\begin{aligned}
\min_{\rho}~~&0.5\sum_{i=1}^m(\rho_i - r_i)^2\; \text{s.t.}\; \rho_i \ge \rho_1 + st_i, \quad i \geq 2,
\end{aligned}\]
which requires a specialized subroutine.
The approach is summarized in
Algorithm~\ref{alg:pg_w}.

\begin{algorithm}[H]
\caption{Source Finding Algorithm}
\label{alg:pg_w}
\begin{algorithmic}[1]
\State{\bfseries Input:} $\{x_i\}$, $s$
\State{\bfseries Initialize:} $k = 0$, $x_s^0 = x_1 $, $w_i^0 = x_s^0 - x_i$

\While{not converged}
\Let{$x_s^{k+1}$}{$ \frac{1}{m} \sum_{i=1}^m (x_i + w_i^{k}) $}
\Let{$w^{k+1}$}{$\proj_{\Omega}(\{x_s^{k+1} - x_i\})$}
\Let{$k$}{$k+1$}
\EndWhile
\State{\bfseries Output:} $x_s^k$
\end{algorithmic}
\end{algorithm}
%Once we have the optimal solution $\overline w$ from Algorithm~\ref{alg:pg_w}, 
%we apply~\eqref{eq:xw} to get the the final $x_s(\overline w)$. 
Algorithm~\ref{alg:pg_w} terminates when the function value or step-size is less than a specified tolerance, 
or if we hit an iteration cap of 200. It is equivalent to proximal gradient descent on the 
value function for~\eqref{eq:obj}:
\[
\widetilde f(w)  = \min_{x_s} f(x,w).
\]
%The step-size criterion is thus equivalent to finding a near-stationary point to~\eqref{eq:obj}.
See~\cite{zheng2018fast} for an analysis of such algorithms, including rates of convergence.
\vskip 8pt
{\bf Robust modification.}
The data collection process by a diagnostic catheter is inherently noisy, and some trial points give anomalous timing data.  
These anomalies then give incorrect information about the feasibility region~\eqref{eq:constraints}. 
To make the method more robust, we detect and remove potential outliers in the course 
of solving each optimization problem~\eqref{eq:obj}. The outliers naturally give constraints that are very hard to satisfy. We introduce a vector $\tau$ to indicate which constraints are easy to fit, and which are difficult. The few constraints that are the least consistent with the remaining data are likely outliers.  This idea can be traced back to least trimmed squares~\cite{rousseeuw1993alternatives}; see also~\cite{aravkin2016smart} for a survey of modern applications. 

To implement the approach, we remove a small number $h$ of arrival times from consideration in each iteration. 
We sort $\{r_1-st_1, \ldots, r_m-st_m\}$ from least to greatest, and for each index $i$ in the smallest $h$ residuals, we set 
$\tau_i = 0$, while all remaining $\tau_i$ are set to $1$. When we update $x_s$, we modify~\eqref{eq:xupdate} to 
\(
x_s = \frac{1}{m-h}\sum_{i=1}^m \tau_i(x_i + w_i).
\)
This strategy removes the influence of the potential outliers, and decreases the number of touches we need to find the source.

\vskip 8pt
{\bf Online implementation.}
We start with several measurements obtained by a preliminary diagnostic catheter 
with $10$ poles, see Figure~\ref{fig:path}. This sets up the first feasibility problem \eqref{eq:constraints} or~\eqref{eq:coupled}, which
we solve by the reformulation~\eqref{eq:obj} point to sample. 

As we proceed, we consider the last $10$ observations in forming each subsequent 
feasibility problem. If we accumulate a lot of data far away from the source,
the feasibility problem becomes harder to solve; in particular the simple approximation 
of assuming a constant propagation speed $s$ between the potential source $x_s$ and 
all observations is not reasonable.  Algorithm~\ref{alg:pg_w} typically finds a solution (i.e. the next potential $x_s$ given current data) 
within 1-2 iterations for most problems.

\section{Results}
\label{sec:results}
\begin{figure}
\center
\includegraphics[width=0.49\linewidth]{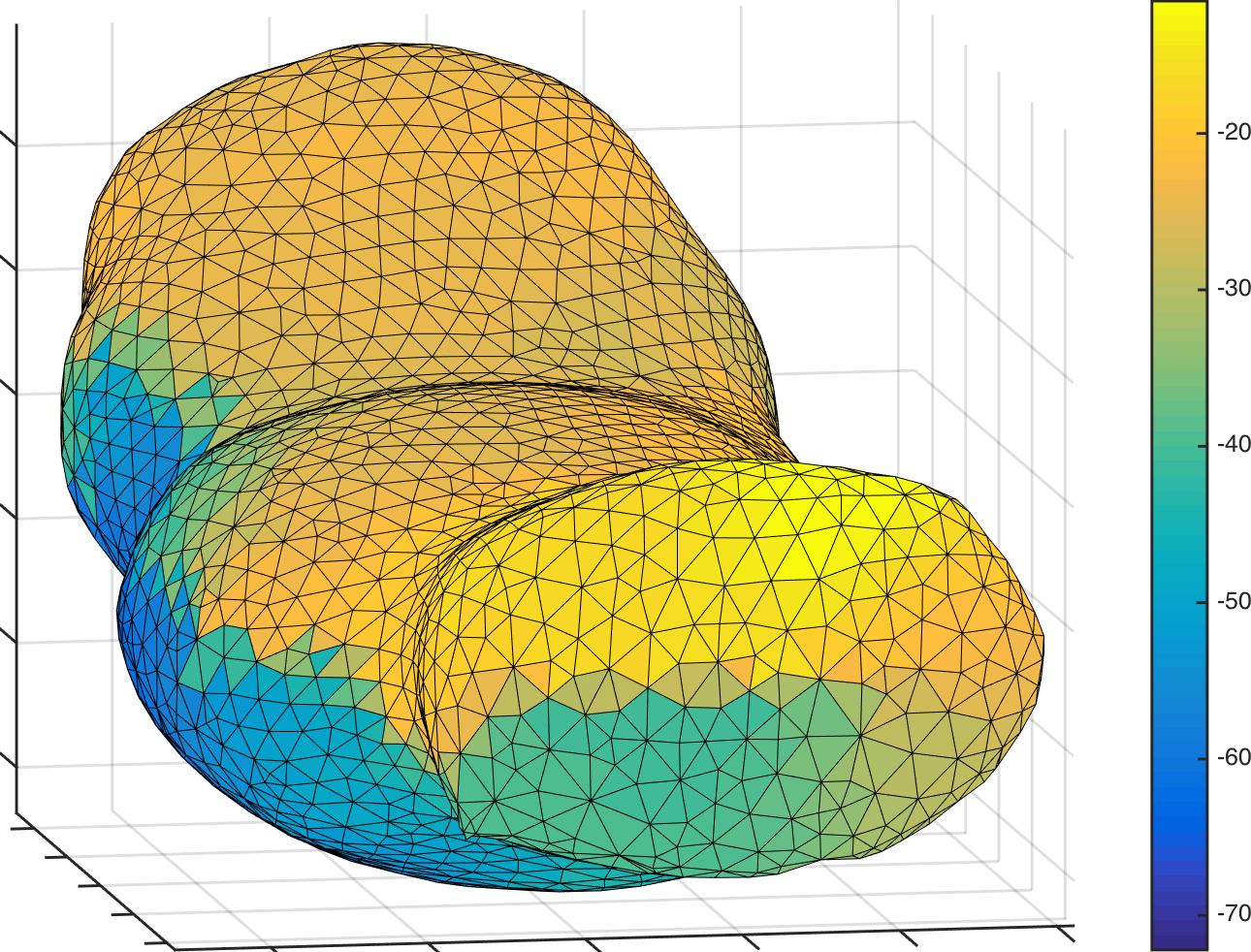}
\includegraphics[width=0.49\linewidth]{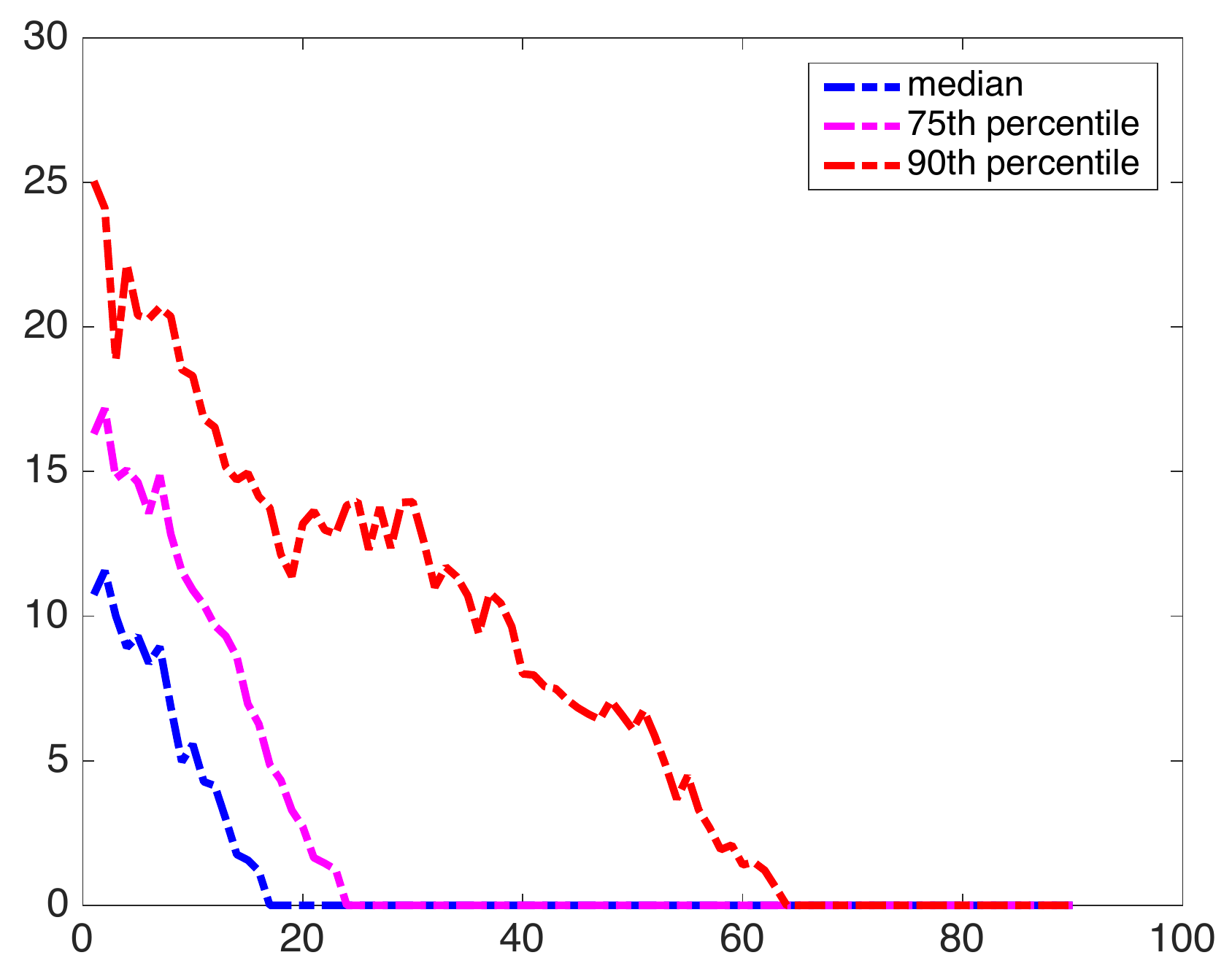}
\caption{\label{fig:timing1} Left: simple ventricle dataset with color-coded arrival times. 
Right: distance to source as a function of touches using simple feasible region~\eqref{eq:constraints}. 
In this simple dataset, the proposed approach quickly finds the source. 
}
\end{figure}
\begin{figure}
\center
\includegraphics[width=0.8\linewidth]{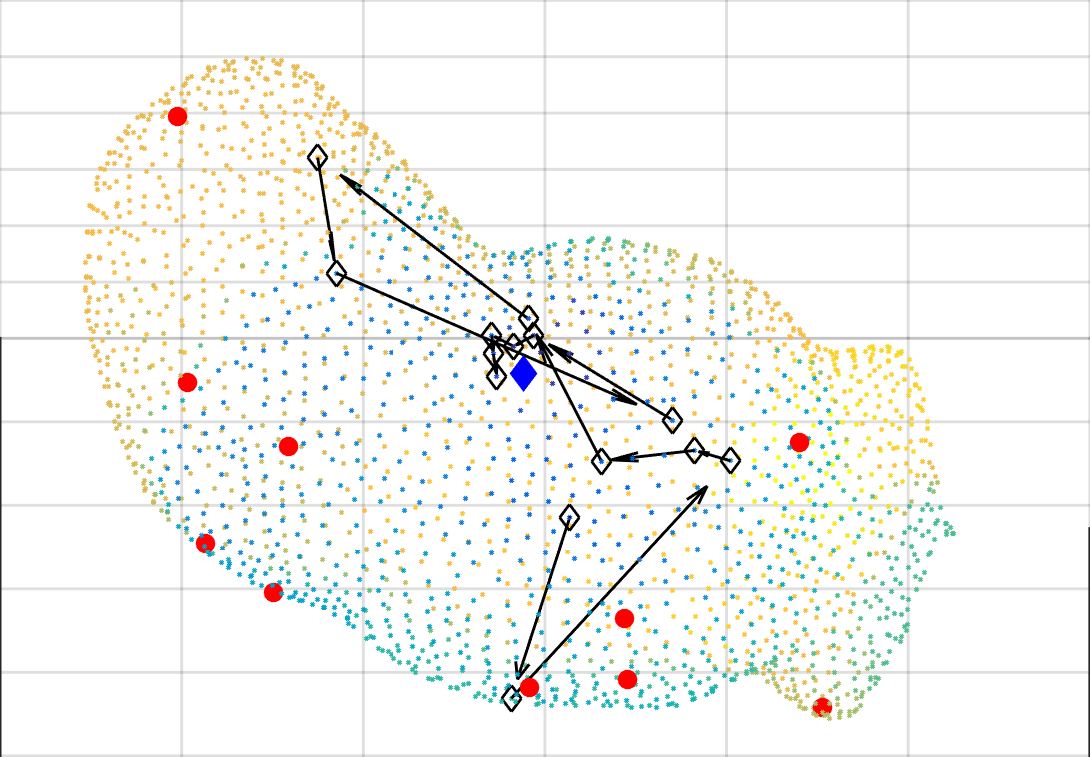}
\caption{\label{fig:path}Path to source for sample run using dataset in Figure~\ref{fig:timing1}. 
Initial placement of diagnostic catheter is shown by red dots; path to source 
is shown using diamond markers.}
\end{figure}
\begin{figure}
\center
\includegraphics[width=0.9\linewidth]{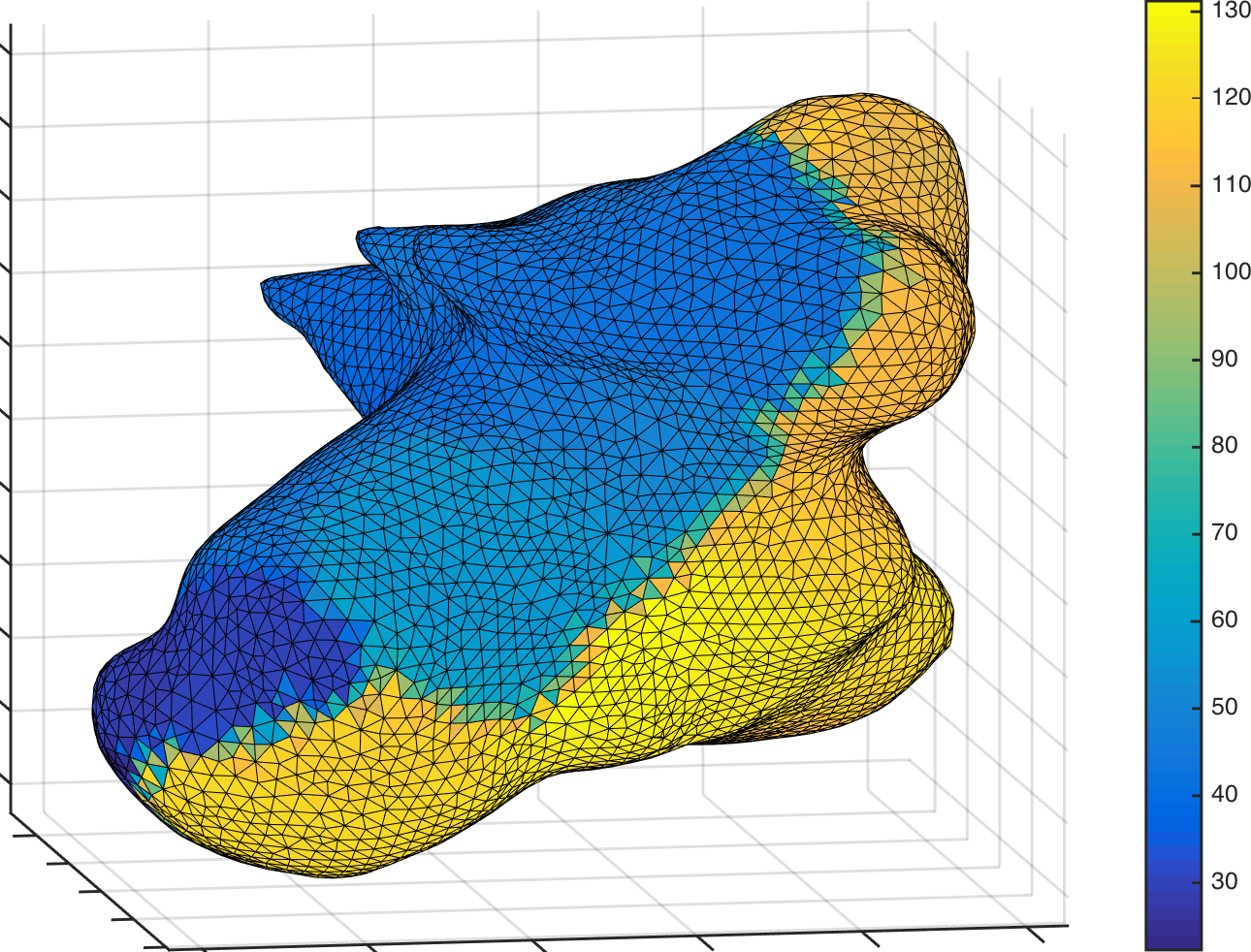}\\
\includegraphics[width=0.49\linewidth]{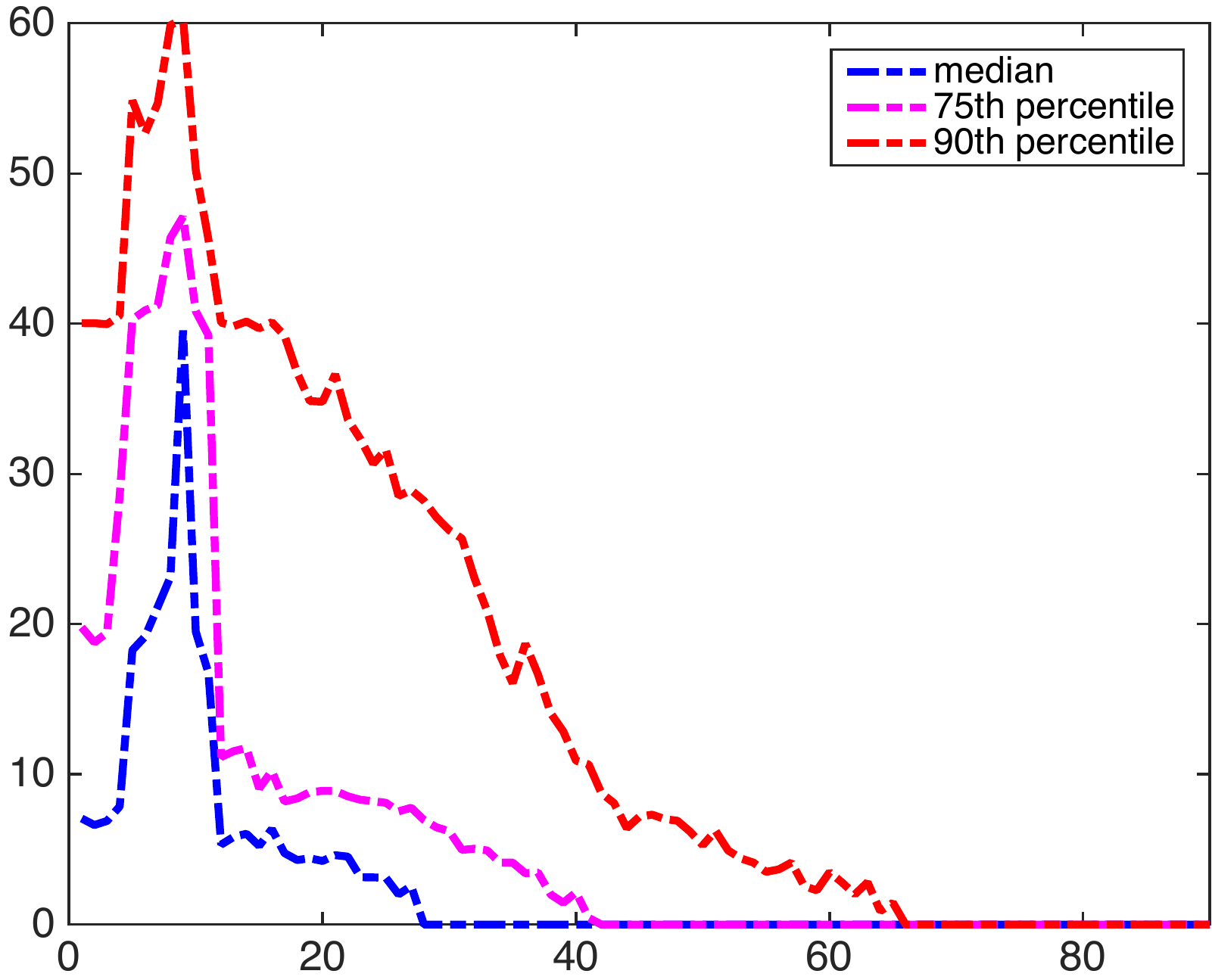}
\includegraphics[width=0.49\linewidth]{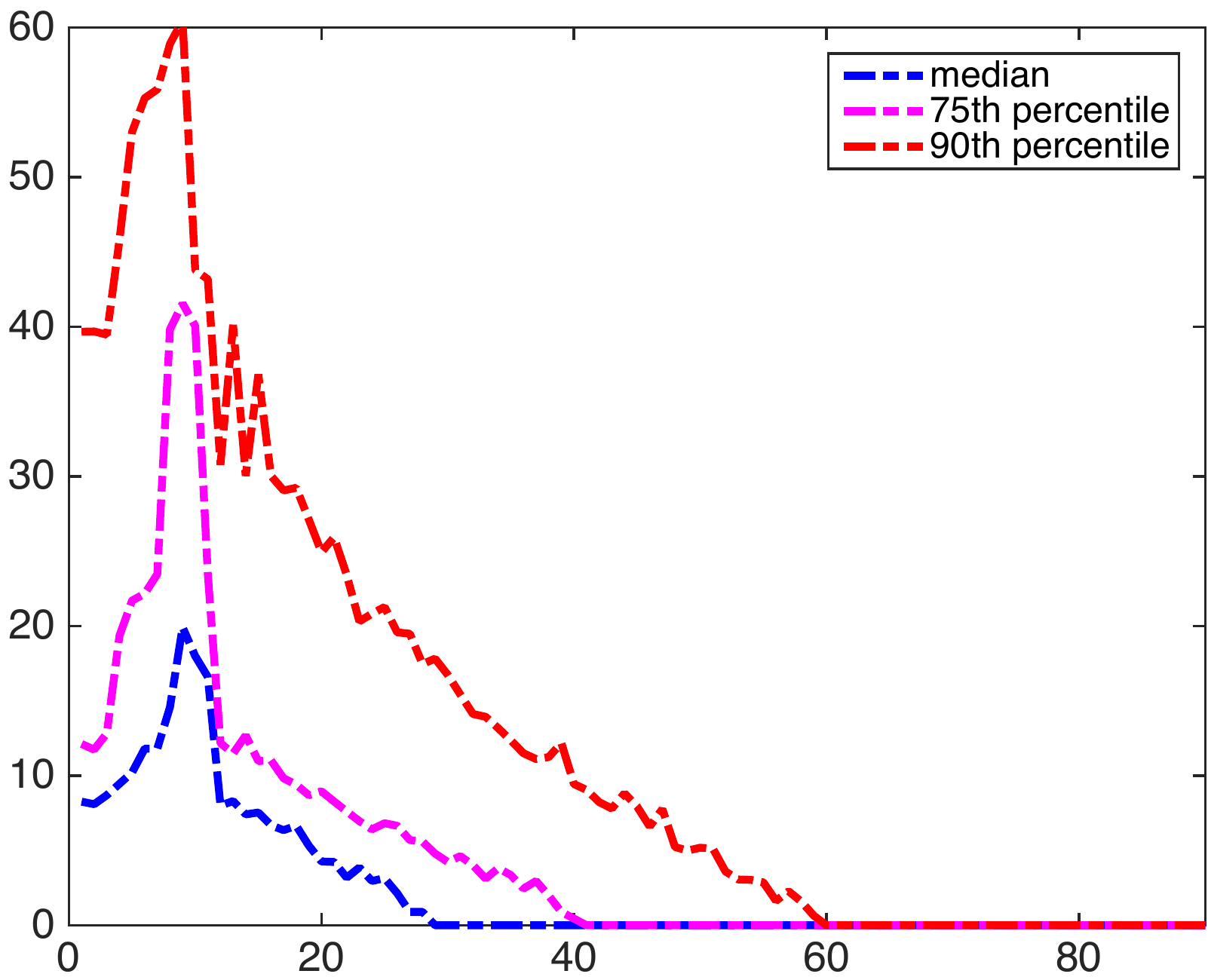}
\caption{\label{fig:timing2}Top panel: ventricle dataset with color-coded arrival times. 
Bottom left: distance to source as a function of touches using simple feasible region~\eqref{eq:constraints}.
Bottom right: distance to source as a function of touches using coupled feasible region~\eqref{eq:coupled}.
Both approaches perform similarly on this example. }
\end{figure}
\begin{figure}
\center
\includegraphics[width=0.9\linewidth]{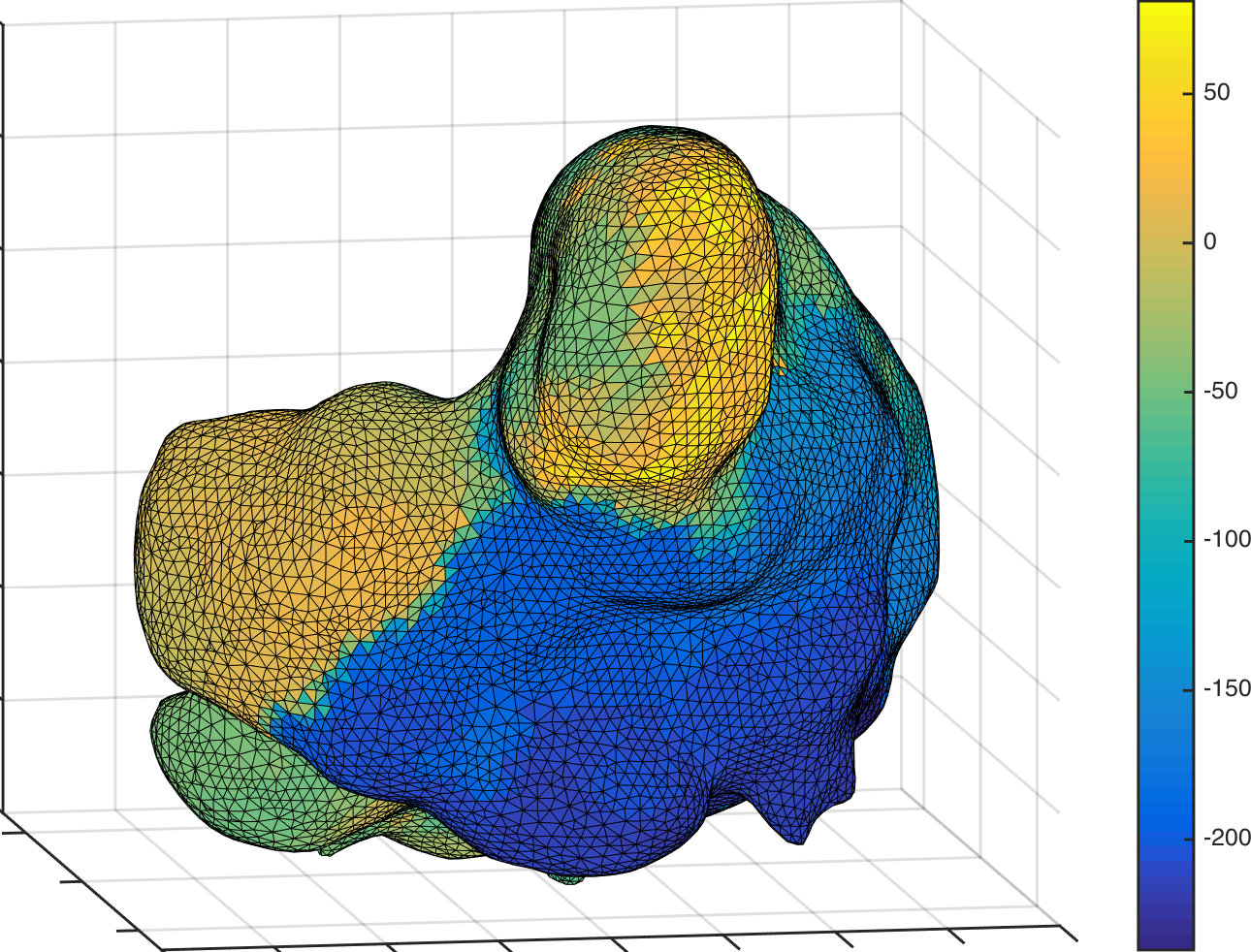}\\
\includegraphics[width=0.49\linewidth]{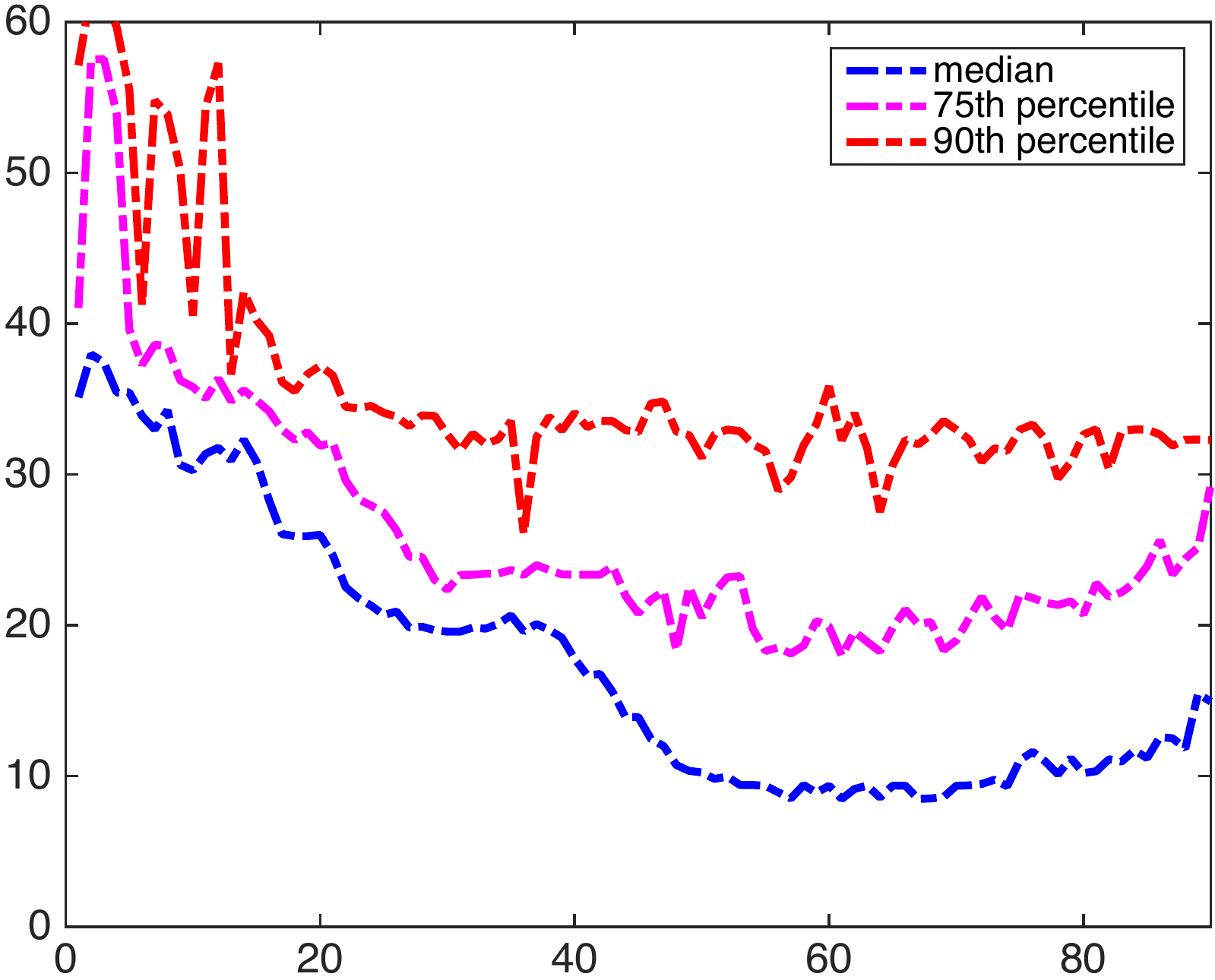}
\includegraphics[width=0.49\linewidth]{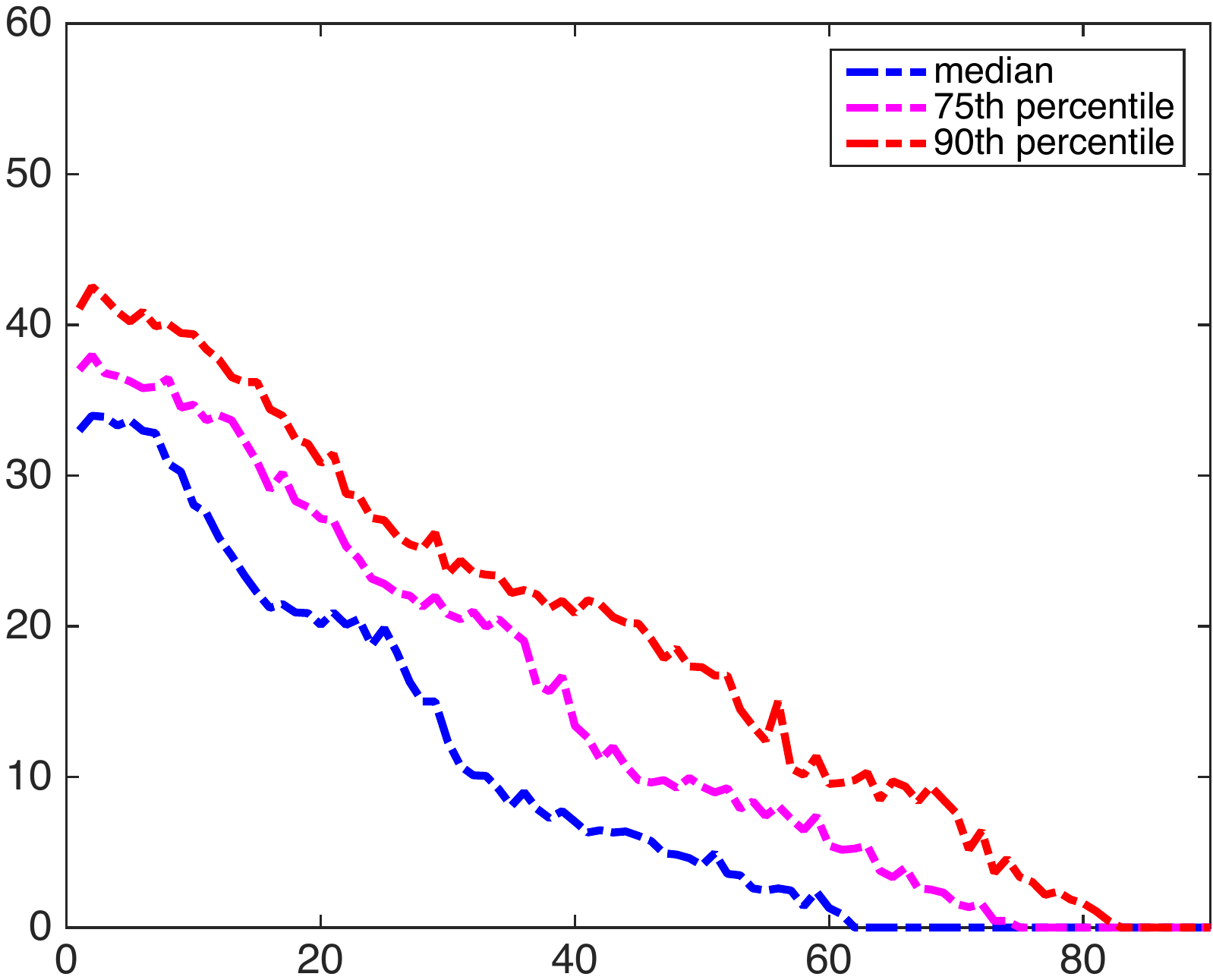}
\caption{\label{fig:timing3} Top panel: atrial dataset with color-coded arrival times. 
Bottom left: distance to source as a function of touches using simple feasible region~\eqref{eq:constraints}.
Bottom right: distance to source as a function of touches using coupled feasible region~\eqref{eq:coupled}.
Accounting for coupling in the signal propagation gives a significant improvement on this challenging dataset. 
}
\end{figure}

We run a simulation using real patient data. Timing and mesh data 
for three patients (two ventricular chambers and one atrium) 
are available from a diagnostic study, see Figures~\ref{fig:timing1}-\ref{fig:timing3}.
The `ground truth' of the arrhythmia source is inferred by the 
earliest arrival time observed during the entire data collection process. We 
start with 10 observations made by a diagnostic catheter, 
and use the online version of the algorithm to locate the source. Monte Carlo sampling is used 
to randomly initialize the initial 10 readings; we track the distance of estimates to source as a function 
of touches, and report the results across the simulations by using median, 75th, and 90th quantiles 
of the distance to source (mm) as a function of touches.
A sample run of the algorithm is shown in Figure~\ref{fig:path}. 
The algorithm proposes the next point 
to sample after each touch. We then sample the 
closest point with data to each $x_s$ proposed by the algorithm; in practice
 the operator can simply attempt to move the catheter to a proposed point to obtain the next measurement.
%\begin{Verbatim}[fontsize = \scriptsize, fontseries=b, fontfamily=courier]
%Touches: 1, curr to src: 6.74e+00
%Touches: 2, curr to src: 6.27e+00
%Touches: 3, curr to src: 5.74e+00
%Touches: 4, curr to src: 6.97e+00
%Touches: 5, curr to src: 4.83e+00
%Touches: 6, curr to src: 5.52e+00
%Touches: 7, curr to src: 4.09e+00
%Touches: 8, curr to src: 4.47e+00
%Touches: 9, curr to src: 3.16e+00
%Touches: 10, curr to src: 1.57e+00
%Touches: 11, curr to src: 1.23e+00
%Touches: 12, curr to src: 1.57e+00
%Found Source
%\end{Verbatim}
The procedure takes 12 touches to locate the true source for this run. For each touch, 
we need 1 or 2 iterations of Algorithm~\ref{alg:pg_w} to solve the feasibility problem. 
The objective values are numerically $0$, which means each $x_s$ we find satisfies~\eqref{eq:constraints}.

Simulation results are plotted for each of the three datasets in Figures~\ref{fig:timing1}-\ref{fig:timing3}. 
For the first two datasets, using the simple feasible region~\eqref{eq:constraints} 
works as well as using the coupled encoding~\eqref{eq:coupled}. For the more complex 
atrial dataset in Figure~\ref{fig:timing3}, adding more geometrical constraints pays off, 
and the results of~\eqref{eq:coupled} are significantly better. In all cases, 
Algorithm~\ref{alg:pg_w} easily solves the nonconvex feasibility problems required to tell 
the operator where to sample next. 

%To test the robustness of the algorithm, we do a simulation study, where we run 100 trials, each initialized 
%with a random initial catheter placement, but using the same data as in Figure~\ref{fig:timing}. We then 
%plot the distance to the source (in millimeters) as a function of the number of touches. 
%\begin{figure}
%\center
%\includegraphics[width=\linewidth]{figs/hard_chamber.jpg}
%\caption{\label{fig:sim} Quantiles of distance to source as a function of the number of touches, using a Monte Carlo simulation with random catheter placement for a deidentified patient dataset. 
%More than 75\% of the runs (magenta curve) are resolved using fewer than 25 touches.}
%\end{figure}
%With no human element, the algorithm finds the source of the arrhythmia within 90 touches in all cases. 
%For 75\% of the trials, we find the exact source within 25 touches, which is a significant threshold for clinicians. 

\section{Conclusion}
\label{sec:conclusion}

We formulated the arrhythmia localization problem as a sequence of nonconvex feasibility problems,  
and developed an efficient algorithm to solve these problems. The approach can accommodate different 
physical models of signal propagation in the heart; we compared two different models in this paper. 
The resulting approach opens a path to a new computationally-guided clinical approach to localize point-source arrhythmias, making online suggestions based on minimal information:  we assume no prior knowledge about 
the heart's anatomy.  The next steps are to develop and test in the clinical setting with existing arrhythmia mapping systems.

\section{Acknowledgements}
We thank Boston Scientific, Inc for providing de-identified Rhythmia® data sets to support this work. 
Dr. Aravkin was supported by the WRF Data Science Professorship. 
\ifCLASSOPTIONcaptionsoff
  \newpage
\fi

\bibliographystyle{abbrv}
\bibliography{references_sasha}

\begin{thebibliography}{1}

\bibitem{aravkin2016smart}
A.~Aravkin and D.~Davis.
\newblock A smart stochastic algorithm for nonconvex optimization with
  applications to robust machine learning.
\newblock {\em arXiv preprint arXiv:1610.01101}, 2016.

\bibitem{li2016douglas}
G.~Li and T.~K. Pong.
\newblock {Douglas--Rachford} splitting for nonconvex optimization with
  application to nonconvex feasibility problems.
\newblock {\em Math. Prog.}, 159(1-2):371--401, 2016.

\bibitem{rousseeuw1993alternatives}
P.~J. Rousseeuw and C.~Croux.
\newblock Alternatives to the median absolute deviation.
\newblock {\em Journal of the American Statistical association},
  88(424):1273--1283, 1993.

\bibitem{sethian1996fast}
J.~A. Sethian.
\newblock A fast marching level set method for monotonically advancing fronts.
\newblock {\em Proceedings of the National Academy of Sciences},
  93(4):1591--1595, 1996.

\bibitem{weber2017novel}
T.~Weber, H.~A. Katus, S.~Sager, and E.~P. Scholz.
\newblock Novel algorithm for accelerated electroanatomic mapping and
  prediction of earliest activation of focal cardiac arrhythmias using
  mathematical optimization.
\newblock {\em Heart rhythm}, 14(6):875--882, 2017.

\bibitem{zheng2018fast}
P.~Zheng and A.~Aravkin.
\newblock Fast methods for nonsmooth nonconvex minimization.
\newblock {\em arXiv preprint arXiv:1802.02654}, 2018.

\end{thebibliography}

\end{document}